# Spin Hall Magnetoresistance Induced by a Non-Equilibrium Proximity Effect

H. Nakayama,[1,][*] M. Althammer,[2,][*] Y.-T. Chen,[3] K. Uchida,[1,4] Y. Kajiwara,[1]
D. Kikuchi,[1,5] T. Ohtani,[1] S. Geprägs,[2] M. Opel,[2] S. Takahashi,[1] R. Gross,[2,6]
G. E. W. Bauer,[1,3,][†] S. T. B. Goennenwein,[2,][‡] and E. Saitoh[1,5,7,8,][§]

[1]*Institute for Materials Research, Tohoku University, Sendai 980-8577, Japan*
[2]*Walther-Meißner-Institut, Bayerische Akademie
der Wissenschaften, 85748 Garching, Germany*
[3]*Kavli Institute of NanoScience, Delft University
of Technology, 2628 CJ Delft, The Netherlands*
[4]*PRESTO, Japan Science and Technology Agency,
Kawaguchi, Saitama 332-0012, Japan*
[5]*WPI Advanced Institute for Materials Research,
Tohoku University, Sendai 980-8577, Japan*
[6]*Physik-Department, Technische Universität München, 85748 Garching, Germany*
[7]*CREST, Japan Science and Technology Agency,
Sanbancho, Tokyo 102-0075, Japan*
[8]*The Advanced Science Research Center,
Japan Atomic Energy Agency, Tokai 319-1195, Japan*

(Dated: October 31, 2012)

## Abstract

We report anisotropic magnetoresistance in Pt|$Y_3Fe_5O_{12}$ bilayers. In spite of $Y_3Fe_5O_{12}$ being a very good electrical insulator, the resistance of the Pt layer reflects its magnetization direction. The effect persists even when a Cu layer is inserted between Pt and $Y_3Fe_5O_{12}$, excluding the contribution of induced equilibrium magnetization at the interface. Instead, we show that the effect originates from concerted actions of the direct and inverse spin Hall effects and therefore call it "spin Hall magnetoresistance."

PACS numbers: 72.25.Ba, 72.25.Mk, 75.47.-m, 75.76.+j

The resistance of a metallic magnet depends on its magnetization direction, a phenomenon called magnetoresistance (MR). Several types of MR, i.e. anisotropic magnetoresistance (AMR) [1], giant magnetoresistance (GMR) [2–4], and tunnel magnetoresistance (TMR) [5–9] are presently indispensable in data storage technology. For these MRs to occur, conduction electrons must pass through the magnet. Here we report the discovery of a fundamentally different MR that is caused by *non-equilibrium* proximity magnetization of a metallic Pt film attached to an electrically insulating magnet $Y_3Fe_5O_{12}$ (YIG). Although the conduction electrons in the Pt film cannot enter the magnetic insulator, the Pt|YIG bilayer resistance reflects the magnetization direction of insulating YIG.

Spin transport and charge transport phenomena are interconnected. For example, the spin Hall effect (SHE) refers to conversion of an electric current into a transverse spin current, i.e. a net flow of electron magnetic moments, due to the spin-orbit interaction (SOI). The conversion efficiency of the SHE is enhanced in heavy metals such as Pt in which the SOI is very strong [Fig. 1(a)]. The reciprocal of the SHE is the inverse spin Hall effect (ISHE), i.e. the conversion of an injected spin current into a transverse electric current or voltage [Fig. 1(c)]. Here the directions of electric-current flow, $\mathbf{J}_e$, spin-current flow, $\mathbf{J}_s$, and spin-current polarization, $\boldsymbol{\sigma}$, are at right angles to one another [10–14].

The SHE generates spin currents and spin accumulations. On the other hand, the ISHE has become useful for detecting spin currents and spin-based electric power generation [10–14]. Here a question arises: Is it possible that SHE and ISHE operate simultaneously? Based on our recent understanding of interfacial spin mixing at the interface between a magnetic insulator and a metal [15–17], we can now answer this question affirmatively.

Consider a free standing metallic thin film exhibiting strong SOI, e.g. Pt. An electric current along the film plane is applied to the Pt film. This $\mathbf{J}_e$ induces a spin current $\mathbf{J}_s$ due to the SHE in Pt that travels perpendicular to the film surface [10, 12–14, 18–20] with spin polarization $\boldsymbol{\sigma}$ parallel to the surface, as shown in Fig. 1(a). Secondly, at the film surface, the spin current is reflected back into the film [see Fig. 1(b)]. In real Pt films, this reflection is responsible for a non-equilibrium spin accumulation near the surface [Fig. 1(f)] and subsequent spin diffusion [21, 22] as described below. Finally, the ISHE in Pt induces an electric current from the reflected spin current [see Fig. 1(c)], causing an electromotive force along the film plane. This additional electric current due to the combination of SHE and ISHE is always parallel to the original one; electric currents measured in a thin film with spin-orbit interaction inevitably include this additional contribution.

We may now control this process by putting an electrically insulating magnet, e.g. garnet-type YIG, on the Pt surface. This gives rise to interfacial spin mixing between YIG and Pt [15, 16], i.e., to the spin-angular-momentum exchange between magnetization **M** in YIG and conduction-electron spin polarization $\boldsymbol{\sigma}$ in Pt. Spin flip scattering is activated when $\boldsymbol{\sigma}$ and **M** are not collinear, as shown in Figs. 1(d) and 1(e). A part of the spin current is then absorbed by the magnetization as a spin-transfer torque [23–25] even at an interface to a magnetic insulator [17] and the spin-current reflection is suppressed. This absorption is maximized when **M** is perpendicular to $\boldsymbol{\sigma}$ and zero when **M** is parallel to $\boldsymbol{\sigma}$ [26]. Therefore, the conductivity enhancement due to SHE and ISHE is expected to be maximized (minimized) when **M** is perpendicular (parallel) to $\mathbf{J}_e$, since $\mathbf{J}_e$ is perpendicular to $\boldsymbol{\sigma}$. The Pt-film resistance is therefore affected by the magnetization direction in YIG, giving rise to the spin Hall magnetoresistance (SMR). Since the SMR in Pt is caused by non-equilibrium spin currents and appears only in the vicinity of the attached YIG on the scale of the spin diffusion length [22], Pt films with thicknesses on comparable scale are necessary. We prepared a 12-nm-thick Pt film on a single-crystalline (111) YIG film [Fig. 2(a)]. YIG is a ferrimagnetic insulator with a large charge gap of $\sim 2.7$ eV [15, 17]; its resistivity is larger than that of air, exceeding $10^{12}$ $\Omega$cm. We measured the resistance $R$ of the Pt film at room temperature.

Figure 2(b) shows the observed resistance change $\Delta R(H) = R(H) - R(H = 0)$ in the Pt film as a function of the amplitude of the magnetic field $H$. In the experiment, the external magnetic field **H** was applied in the Pt film plane, perpendicular to the electric current direction. In the present field range, the resistance of Pt depends on $H$ only very weakly. However, as shown in Fig. 2(b), the Pt|YIG bilayer surprisingly exhibits a clear resistance change for $|\mathbf{H}| < 20$ Oe. The resistance decreases when increasing $|\mathbf{H}|$ from $H = 0$ with a small hysteresis. In contrast, outside this field range, viz. $|\mathbf{H}| > 20$ Oe, the resistance is almost constant. The field range in which the resistance change appears coincides with the remagnetization process of the YIG layer such that $R(H)$ has a maximum at the coercive fields of YIG. Figure 2(c) shows the $H$ dependence of the in-plane magnetization of the YIG layer; the magnetization change is saturated for $|\mathbf{H}| > 20$ Oe, in agreement with the observed magnetoresistance. The longitudinal magnetoresistance MR $= \Delta R(H)/R(H = 0)$ in the Pt film clearly reflects the remagnetization of the YIG film.

Since Pt is near the Stoner ferromagnetic instability, ferromagnetism induced in the Pt layer by the equilibrium proximity to YIG appears possible and could give rise to the AMR [27–30]. In fact, Pt atoms very close to the interface in Pt|Fe films are known to develop a finite magnetic moment due to a static proximity effect [28]. However, by systematic measurements on a number of reference structures, we are able to prove that such a proximity effect cannot be invoked to explain our observations in Pt|YIG. First, the SMR effect appears even when a 6-nm-thick Cu layer is inserted between the Pt and the YIG layers as shown in Fig. 3(b); Cu is very far from the Stoner instability and the non-local exchange force does not reach over such a thickness. Cu has a long (several hundred nanometers) spin diffusion length [21, 22] and a very small SHE, viz. weak spin-orbit interaction, and carries a spin current over long distances. The observation of the MR signal in Pt|Cu|YIG clearly shows that a magnetized Pt layer cannot explain the observed MR. The reduced MR ratio in Fig. 3(b) relative to Fig. 3(a) is caused by the short-circuit current path through the highly-conductive Cu spacer [31]. Before sputtering the Pt layer of this sample, we confirmed that at least more than 95 % of the YIG surface is covered with Cu film using energy dispersive X-ray analysis for the whole surface. For confirmation, we furthermore checked that the MR signal disappears by replacing the Cu layer with an insulating $SiO_2$ layer as shown in Fig. 3(c), where $SiO_2$ is a non-magnetic insulator allowing no spin current to pass. The MR signal also disappears in a 6-nm-thick single layer Cu film on YIG [Fig. 3(d)] in which the spin-orbit interaction is very weak [15, 16], indicating the crucial role of the spin-orbit interaction, or the SHE, while electromagnetic artifacts can also be excluded as origin of the MR.

Finally, we found that the magnetoresistance in the present system exhibits a magnetic field orientation dependence that is very different from the AMR, but consistent with the SMR scenario sketched above (cf. Fig. 4), confirming again the irrelevance of the AMR in a magnetized Pt layer. The AMR and SMR critically differ in their angle dependence: the AMR is governed by the angle of the applied current to the magnetization direction [1], $\alpha_{cM}$, while the SMR depends on the angle of the spin accumulation induced by the current with the magnetization, $\alpha_{\sigma M}$. This difference becomes manifest when the magnetic field direction is swept from the direction parallel to the electric current direction to the direction normal to the film surface [Fig. 4(k), $\alpha = 0$, $\gamma = +90° \to 0$]. During this field-direction scan, $\alpha_{\sigma M}$ keeps constant while $\alpha_{cM}$ varies from 90° to 0, and according to the theory of the SMR, and in contrast to the AMR, the resistance should not change. This unusual behavior offers a key test of the SMR scenario. Figure 4(e) shows the MR of a Pt|YIG sample measured with changing the field direction from $\gamma = -90° \to 290°$ at $\alpha = 0$ [see Fig. 4(k)]. Since the magnetic field intensity is fixed at 12 kOe, far above the magnetization saturation field ($\sim 1.7$ kOe), the magnetization is always aligned with the external magnetic field direction. Clearly, the $\Delta R$ signal disappears in this magnetic field orientation scan, in striking contrast to the other field-direction scans [cf. Figs. 4(c)−4(f)]. This behavior is observed not only in the present sample (sample 1) but also universally in our qualitatively different Pt|YIG samples as exemplified for sample 2 [31] in Fig. 4(i). The disappearance of $\Delta R$ is a unique feature of the SMR and cannot be explained by the AMR. We thus conclude that the AMR of a conventional, equilibrium proximity spin polarization in Pt can be ruled out as an explanation for the magnetoresistance observed in experiment.

We introduced the SMR in a simple ballistic picture of spin currents reflected at the interfaces. For quantitative modeling it is necessary to invoke the diffusive nature of transport as well as spin dissipation in the metallic film. Considering a thin Pt film in the $xy$-plane with an electric current applied along the $x$-direction, the SHE generates a spin current flowing in the $z$-direction with the spin polarization along $y$-direction, thereby building up spin accumulations at the Pt|YIG and vacuum|Pt interfaces. Their gradients induce diffusive counter spin currents such that the total (net) spin current is continuous at the Pt|YIG interface and vanishes at the vacuum|Pt surface. The interface spin current depends on the relative direction of the magnetization with respect to the spin accumulation direction according to $(G_r/e)\mathbf{m}\times(\mathbf{m}\times\boldsymbol{\mu}_s)$, where $G_r$ is the interface spin-mixing conductance, $\mathbf{m}$ is the magnetization direction, and $\boldsymbol{\mu}_s$ is the spin accumulation vector at the interface [26]. When $\mathbf{m} \parallel \boldsymbol{\mu}_s$, the interface spin current vanishes (just as at the vacuum interface). However, when the magnetization is rotated by 90° (to any perpendicular direction), the accumulated spins are partially absorbed at the interface and dissipated as a spin-transfer torque to the magnetization, thereby modulating the spin-current distribution in Pt. Since the current-induced spin accumulation is polarized along the $y$-direction, the polarization direction of the modulated spin current flowing along $z$ varies as $\propto \mathbf{m}\times(\mathbf{m}\times\hat{\mathbf{y}})$. This in turn modulates the longitudinal (applied) electric current as $\propto \hat{\mathbf{y}}\cdot[\mathbf{m}\times(\mathbf{m}\times\hat{\mathbf{y}})] = m_y^2 - 1$ and induces a transverse electric current $\propto \hat{\mathbf{x}}\cdot[\mathbf{m}\times(\mathbf{m}\times\hat{\mathbf{y}})] = m_x m_y$ in the $y$-direction due to the ISHE, where $m_x$ and $m_y$ are the Cartesian components of $\mathbf{m}$. The prefactors of these dependencies can be computed by spin diffusion theory [14] and quantum mechanical boundary conditions in terms of the spin-mixing conductance [26], thereby fully explaining the observed SMR in Pt|YIG [31]. The SMR resistivity change can hence be formulated as

$$\rho = \rho_0 - \Delta\rho m_y^2, \ \rho_{\text{trans}} = \Delta\rho m_x m_y. \qquad (1)$$

This is very different from the AMR phenomenology of polycrystalline conductive ferromagnets [1]

$$\rho = \rho_\perp + \Delta\rho_A m_x^2, \ \rho_{\text{trans}} = \Delta\rho_A m_x m_y. \qquad (2)$$

In both expressions the resistivity $\rho$ is measured along the direction of the electric current flow $\mathbf{J}_e$ (along $x$-direction, cf. Fig. 4), while $\rho_{\text{trans}}$ is the resistivity component recorded in the sample plane perpendicular to $\mathbf{J}_e$ (along $y$-direction) which typically appears in the magnetoresistive properties of ferromagnets [1]. $\rho_0$ is a constant resistivity offset, $\Delta\rho$ and $\Delta\rho_A$ ($= \rho_\parallel - \rho_\perp$) are the magnitude of the resistivity change as a function of the magnetization orientation, $\rho_\parallel$ and $\rho_\perp$ are the resistivities for magnetizations aligned along and perpendicular to $\mathbf{J}_e$, respectively. In Figs. 4(a) and 4(b), we show the evolution of the MR and $\text{MR}_{\text{trans}} = \rho_{\text{trans}}(H)/\rho(H = 0)$ in sample 1 as a function of $H$, applied at different angles $\alpha$. To quantitatively evaluate this dependence, we show the evolution of the MR and $\text{MR}_{\text{trans}}$ as a function of $\alpha$ in Figs. 4(c) and 4(d), respectively (symbols). The MR for rotations of the magnetization in the plane perpendicular to the $y$-direction (angle $\gamma$) and perpendicular to the $x$-direction (angle $\beta$) are summarized in Figs. 4(e) and 4(f), while Figs. 4(i) and 4(j) show corresponding transport data for the sample 2. The behavior of the electric resistance expected from the AMR according to Eq. (2) is shown as blue curves in these panels, while the SMR predicted by Eq. (1) is depicted by red curves. The out-of-plane rotation data are consistently described in terms of the SMR; the angle-dependent MR data thus show that the MR observed in experiment indeed is due to the SMR effect. For a 12-nm-thick Pt film with the resistivity $8.6 \times 10^{-7}$ $\Omega$m the theory sketched above agrees with the experimental $\Delta\rho$ for the spin Hall angle $\theta_{\text{SH}} = 0.04$, the spin-flip diffusion length of $\lambda = 2.4$ nm, and the spin-mixing conductance [17] of $G_r = 5 \times 10^{14}$ $\Omega^{-1}\text{m}^{-2}$.

The spin Hall magnetoresistance (SMR) is a non-equilibrium proximity effect: the resistance of the metal film depends on the magnetic properties of the adjacent, electrically insulating ferromagnet. The SMR is not caused by a statically induced magnetization and is qualitatively different from conventional magnetoresistance effects, such as AMR, GMR, and TMR, where an electric current must flow through the magnetic layers. The SMR en-ables remote electrical sensing of the magnetization direction in a magnetic insulator. This also implies that the SMR makes the integration of insulating ferromagnets into electronic circuits possible, thereby avoiding current induced deterioration of magnets due to, e.g., electromigration or heating. Finally, the SMR allows studying and quantifying spin Hall effects in paramagnetic metals as well as spin transfer to magnetic insulators via simple d.c. magnetoresistance measurements. We anticipate that SMR will develop into a standard technique in the nascent field of insulator spintronics.

The work at Tohoku University was supported by CREST-JST "Creation of Nanosystems with Novel Functions through Process Integration," Japan, PRESTO-JST "Phase Interfaces for Highly Efficient Energy Utilization," Japan, Grant-in-Aid for JSPS Fellows from JSPS, Japan, a Grant-in-Aid for Scientific Research A (24244051) from MEXT, Japan, a Grant-in-Aid for Research Activity Start-up (24860003) from MEXT, Japan, LC-IMR of Tohoku University, and the Murata Science Foundation. The work at the Walther-Meissner-Institut and TU Delft was supported by the Deutsche Forschungsgemeinschaft (DFG) through priority programme SPP 1538 "Spin-Caloric Transport," project GO 944/4. G.E.W.B. acknowledges support from the Dutch FOM Foundation and EC-Project MACALO.

* These authors contributed equally to this work.

† Electronic address: g.e.w.bauer@imr.tohoku.ac.jp

‡ Electronic address: goennenwein@wmi.badw.de

§ Electronic address: saitoheiji@imr.tohoku.ac.jp

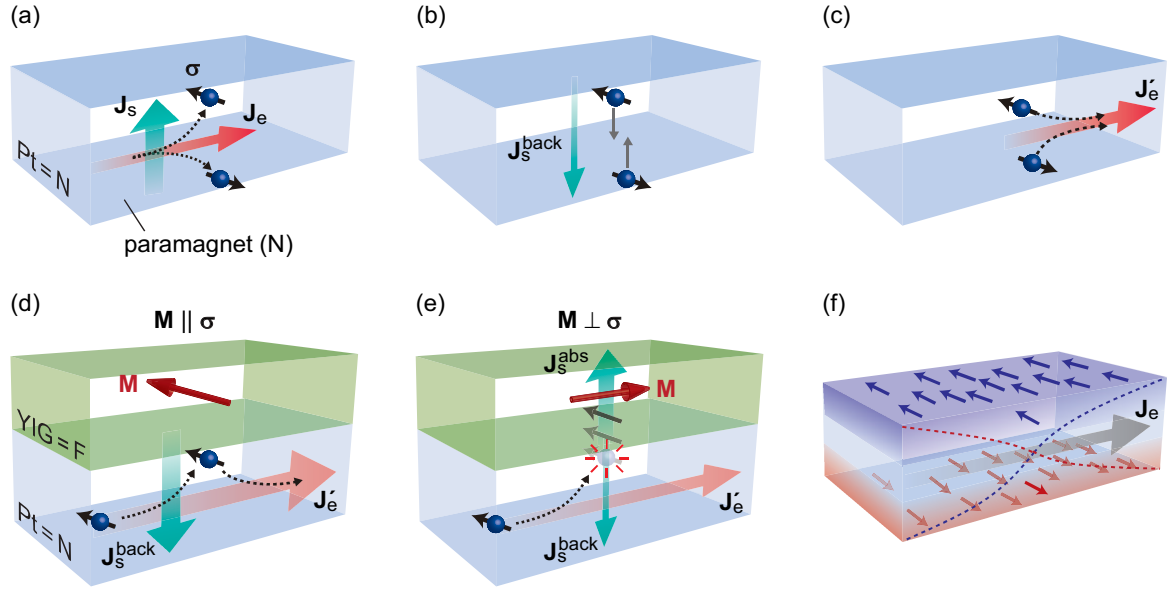

FIG. 1: (a)-(c) Illustrations of the magnetic control of the conductivity due to the direct and inverse spin Hall effects (SHE and ISHE) in a paramagnetic thin film metal (N) with strong spin-orbit interaction attached to a ferromagnetic insulator (F). (d)-(e), Illustrations of the geometric relation between flow of electron and accumulated spins in N = Pt and the magnetization in the magnetic insulator F = YIG. (f) Schematic illustration of the spin accumulation generated by non-equilibrium proximity due to the SHE in N. At the interfaces of N, the spin accumulation is formed depending on its spin polarization direction. Dashed curves in N show the electron motions with different spin polarization directions; the blue (red) arrows move to the upper (under) side.

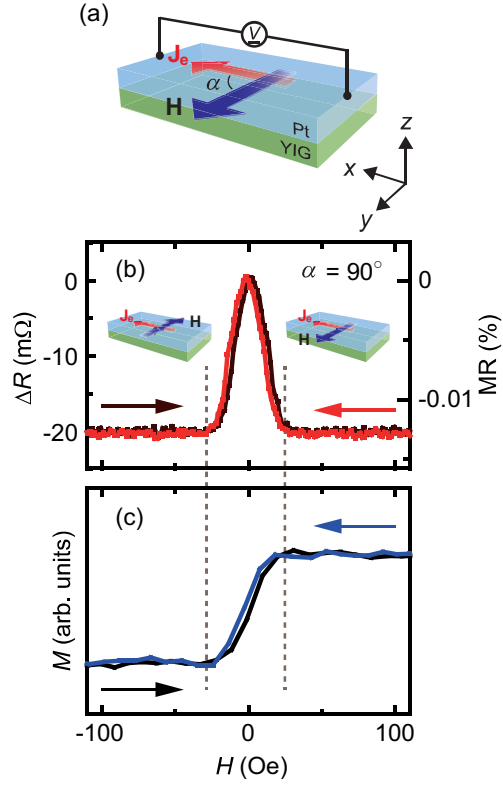

FIG. 2: (a) Illustration of the experimental set up. The sample is a Pt|YIG bilayer film composed of a 1.3-$\mu$m-thick magnetic insulator YIG layer and a 12-nm-thick Pt layer. The electric resistance is measured by the four probe method. Here, $\mathbf{J}_\mathrm{e}$, $\mathbf{H}$, and $\alpha$ represent the electric current in the Pt layer, the external magnetic field, and the relative in-plane angle between $\mathbf{J}_\mathrm{e}$ and $\mathbf{H}$, respectively. (b) Magnetoresistance (MR) $\Delta R$ for $\alpha = 90°$. (c) Magnetization $M$ of a plain YIG film at 300 K.

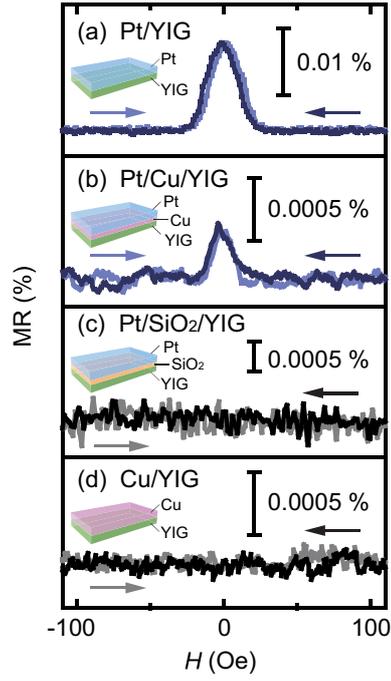

FIG. 3: Control MR experiments on (a) Pt|YIG, (b) Pt|Cu|YIG, (c) Pt|SiO$_2$|YIG, and (d) Cu|YIG composite films, respectively. The in-plane external magnetic field is applied perpendicular to the electric current, $\alpha = 90°$. The insets sketch the different samples.

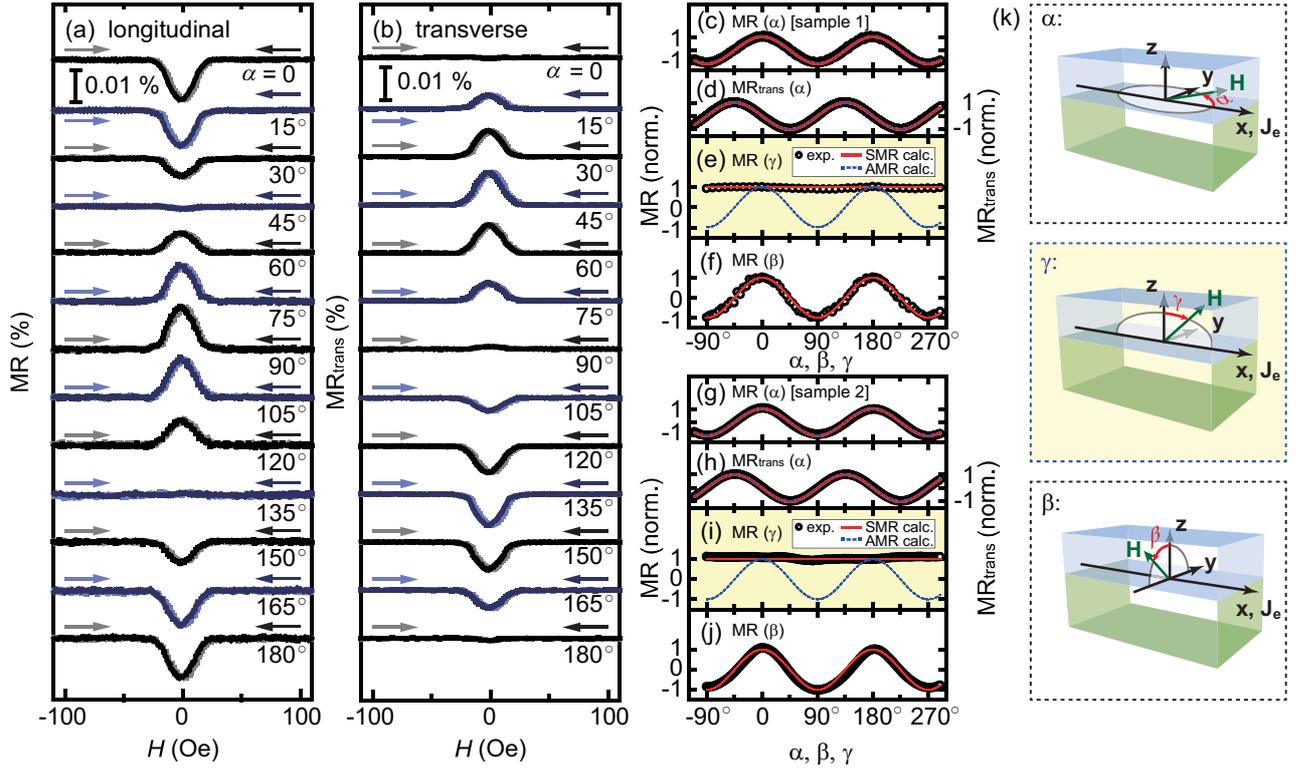

FIG. 4: (a)-(b) Longitudinal and transverse MR in Pt|YIG films as a function of in-plane angle $\alpha$. (c)-(j) $\alpha$, $\beta$, and $\gamma$ dependence of the normalized longitudinal and transverse MR in two different samples, where the angles $\alpha$, $\beta$, and $\gamma$ are defined in (k). The red and blue curves show MR expected according to the SMR model and the AMR model, respectively. (c) and (g) show $\alpha$ dependence of the longitudinal MR effect, (d) and (h) show $\alpha$ dependence of the transverse MR effect, (e) and (i) show $\gamma$ dependence of the longitudinal MR effect, and (f) and (j) show $\beta$ dependence of the longitudinal MR effect, respectively.

# Spin Hall Magnetoresistance Induced by a Non-Equilibrium Proximity Effect

H. Nakayama[*], M. Althammer[*], Y.-T. Chen, K. Uchida, Y. Kajiwara, D. Kikuchi,
T. Ohtani, S. Geprägs, M. Opel, S. Takahashi, R. Gross, G. E. W. Bauer,
S. T. B. Goennenwein, and E. Saitoh

## SUPPLEMENTAL MATERIAL

**S1. Materials and Methods**

The sample 1 consists of a single-crystalline film of the ferrimagnetic insulator $Y_3Fe_5O_{12}$ (YIG) (111) with the thickness of 1.3 μm grown on an insulator $Gd_3Ga_5O_{12}$ (111) single-crystalline substrate by liquid phase epitaxy. The film was deposited under a $PbO$-$B_2O_3$ flux at around 1,200 K. Subsequently, 12-nm-thick Pt layer was sputtered on the top of the YIG layer and patterned into a Hall bar. The resistivity of the Pt film is typically $\rho_0 \sim 8.6 \times 10^{-7}$ Ωm, with the in-plane MR ratio $[\rho(\alpha = 0) - \rho(\alpha = 90°)]/\rho_0 \sim 2.5 \times 10^{-4}$. The sample 2 (Figs. 4 g-j in the main text) consists of a 20-nm-thick YIG film epitaxially grown by pulsed laser deposition from a polycrystalline target onto a single-crystalline, (111)-oriented $Gd_3Ga_5O_{12}$ substrate. The YIG growth took place in an oxygen atmosphere at a pressure of 25 μbar, using laser pulses with an energy density of 2 J/cm$^2$ at the target, and at a substrate temperature of 550°C. The YIG thin film was then covered in-situ, without breaking the vacuum, with a 7-nm-thick Pt film obtained from electron beam evaporation. After removing the sample from the growth chamber, the thin film heterostructure was patterned into Hall bar mesa structures using photolithography and Ar ion beam milling. The resistivity of the Pt film in the sample 2 is $\rho_0 \sim 4.1 \times 10^{-7}$ Ωm, with the in-plane MR ratio $\sim 6 \times 10^{-4}$. The Cu layers (Figs. 3b and 3d in the main text) were fabricated by electron beam evaporation in ultra-high vacuum. The $SiO_2$ spacer layer of the control sample (see Fig. 3c in the main text) was fabricated by sputtering.

All magneto-transport data were taken at room temperature, using current-bias, 4-wire voltage measurements. The magnetic field orientation dependent data were taken with a magnetic field large enough to align the magnetization vector along the external field. The magnetization curve for the plain YIG film as shown in Fig. 2c in the main text was measured by the magneto optic Kerr effect.

---

[*] These authors contributed equally to this work.

## S2. Theory

Here we formulate the electron transport theory for a bilayer system consisting of a normal metal layer (N) with the thickness $d_N$ in contact with an insulating ferromagnet (IF) (Fig. S1). The normal metal is treated by the spin-diffusion theory in the presence of spin-orbit interaction [S1] with boundary conditions at the N|IF interface from the microscopic scattering theory [S2]. We focus on the longitudinal and transverse resistances of the Pt|YIG systems as a function of the magnetization direction presented in the main text, i.e. the spin Hall magnetoresistance (SMR).

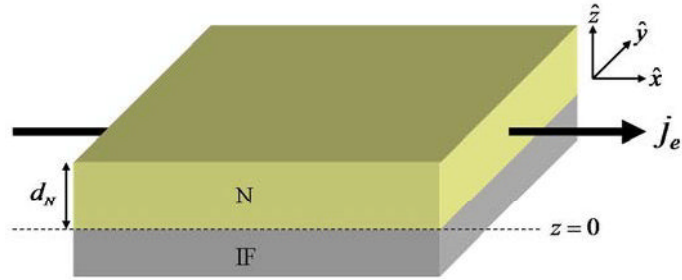

**Figure S1 | The normal metal (N)|electrically insulating ferromagnet (IF) bilayer, where $j_e$ is the electric current density in Pt.**

The basic linear response relation (Ohms's Law) in the normal metal can be described by the following matrix relation between the driving forces, i.e. the electric field $\vec{E} = E_x \hat{x}$ and the gradient of the spin accumulation $\vec{\mu}_s$, and the induced currents:

$$\begin{pmatrix} \vec{j}_e \\ \vec{j}_{sx} \\ \vec{j}_{sy} \\ \vec{j}_{sz} \end{pmatrix} = \sigma_N \begin{pmatrix} 1 & \theta_{SH}\hat{x}\times & \theta_{SH}\hat{y}\times & \theta_{SH}\hat{z}\times \\ \theta_{SH}\hat{x}\times & 1 & 0 & 0 \\ \theta_{SH}\hat{y}\times & 0 & 1 & 0 \\ \theta_{SH}\hat{z}\times & 0 & 0 & 1 \end{pmatrix} \begin{pmatrix} \vec{E} \\ -\vec{\nabla}\mu_{sx}/(2e) \\ -\vec{\nabla}\mu_{sy}/(2e) \\ -\vec{\nabla}\mu_{sz}/(2e) \end{pmatrix}. \quad (S1)$$

Here $\vec{j}_e$ is the electric current and $\vec{j}_{si}$ is the spin current polarized in the $\hat{i}$-direction. Both vectors point in the particle flow direction and all currents are expressed in Ampere. The spin accumulation $\mu_{si}$ is polarized in the $\hat{i}$-direction. The vector product $\times$ operates on the driving forces. The phenomenological parameters are $\sigma_N$, the electric conductivity and $\theta_{SH}$, the "spin Hall angle".

The spin accumulation $\vec{\mu}_s = (\mu_{sx}, \mu_{sy}, \mu_{sz})$ is the solution of the spin-diffusion equation

$$\nabla^2 \vec{\mu}_s = \frac{\vec{\mu}_s}{\lambda^2}, \quad (S2)$$

where $\lambda = \sqrt{D\tau_{sf}}$ is the spin diffusion length expressed in terms of the (charge) diffusion constant $D$ and the spin-flip time $\tau_{sf}$ [S3]. We assume that the system is

translationally invariant in the $x$-$y$ plane, and the solution for the spin diffusion equation reads

$$\mu_{si} = A_i e^{-z/\lambda} + B_i e^{z/\lambda}, \qquad (S3)$$

where $\hat{i} = x, y, z$, and the coefficients $A_i$ and $B_i$ are determined by the boundary conditions at the interfaces $z = d_N$ and $z = 0$.

For $\theta_{SH} \neq 0$ an electric current along $\hat{x}$ generates a spin current in the $\hat{z}$-direction $(\vec{j}_s)_z$, which leads to a spin accumulation build-up at the interfaces that is modulated by the magnetization direction. From equation (S1), the total magnitude of the induced spin current towards the N|IF interface is the sum of the spin Hall current and the spin diffusion current:

$$(\vec{j}_s)_z = -\frac{\sigma_N}{2e} \partial_z \vec{\mu}_s - \theta_{SH} j_e^0 \hat{y}, \qquad (S4)$$

where $j_e^0 = \sigma_N E_x$. At the interface $z = 0$ the spin current is partially absorbed by the ferromagnet as

$$(\vec{j}_s)_z(0) = 2 G_r \hat{m} \times \left( \hat{m} \times \frac{\vec{\mu}_s}{2e} \right), \qquad (S5)$$

where $G_r$ is the real part of the spin mixing interface conductance [S2], while the imaginary part may be disregarded [S4]. Equation (S5) depends on the unit vector of the magnetization direction $\hat{m}$ relative to the spin accumulation and thereby controls the SMR. At the interface with vacuum $z = d_N$, the spin current vanishes

$$(\vec{j}_s)_z(d_N) = -\frac{\sigma_N}{2e} \partial_z \vec{\mu}_s - \theta_{SH} j_e^0 \hat{y} = 0. \qquad (S6)$$

With boundary conditions (S5) and (S6) we can solve the diffusion equation (S2). First we consider the in-plane measurement. For an in-plane magnetization $\hat{m} = (\cos\alpha, \sin\alpha, 0)$

$$\mu_{sx}(z) = 2\theta_{SH} e E_x \lambda \frac{\frac{2\lambda}{\sigma_N} G_r \tanh \frac{d_N}{2\lambda}}{1 + \frac{2\lambda}{\sigma_N} G_r \coth \frac{d_N}{\lambda}} \frac{\cosh \frac{z-d_N}{\lambda}}{\sinh \frac{d_N}{\lambda}} \cos\alpha \sin\alpha, \qquad (S7)$$

$$\mu_{sy}(z) = -2\theta_{SH} e E_x \lambda \left( \frac{\sinh \frac{2z-d_N}{2\lambda}}{\cosh \frac{d_N}{2\lambda}} + \frac{\frac{2\lambda}{\sigma_N} G_r \tanh \frac{d_N}{2\lambda}}{1 + \frac{2\lambda}{\sigma_N} G_r \coth \frac{d_N}{\lambda}} \frac{\cosh \frac{z-d_N}{\lambda}}{\sinh \frac{d_N}{\lambda}} \cos^2\alpha \right), \qquad (S8)$$

and $\mu_{sz} = 0$. Substituting equations (S7) and (S8) into equation (S1), the electric current reads (under closed circuit conditions)

$$\vec{j}_e = j_{ex}\hat{x} + j_{ey}\hat{y}, \tag{S9}$$

in which the longitudinal (along $\hat{x}$) and transverse (along $\hat{y}$) electric currents are

$$j_{ex}(z) = j_e^0 + \theta_{SH}^2 j_e^0 \left( \frac{\cosh\frac{2z-d_N}{2\lambda}}{\cosh\frac{d_N}{2\lambda}} + \frac{\frac{2\lambda}{\sigma_N}G_r \tanh\frac{d_N}{2\lambda}}{1+\frac{2\lambda}{\sigma_N}G_r \coth\frac{d_N}{\lambda}} \frac{\sinh\frac{z-d_N}{\lambda}}{\sinh\frac{d_N}{\lambda}} \cos^2\alpha \right), \tag{S10}$$

$$j_{ey}(z) = \theta_{SH}^2 j_e^0 \frac{\frac{2\lambda}{\sigma_N}G_r \tanh\frac{d_N}{2\lambda}}{1+\frac{2\lambda}{\sigma_N}G_r \coth\frac{d_N}{\lambda}} \frac{\sinh\frac{z-d_N}{\lambda}}{\sinh\frac{d_N}{\lambda}} \cos\alpha \sin\alpha. \tag{S11}$$

The observables are the longitudinal ($\rho$) and transverse ($\rho_{trans}$) resistivities. Averaging the electric currents over the film thickness $z$ and expanding the longitudinal resistivity to leading order in $\theta_{SH}^2$:

$$\rho = \sigma^{-1} = \left(\frac{\overline{j_{ex}}}{E_x}\right)^{-1} \approx \rho_0 + \Delta\rho \cos^2\alpha, \tag{S12}$$

$$\rho_{trans} = -\frac{\sigma_{trans}}{\sigma_N^2} = -\frac{\overline{j_{ey}}/E_x}{\sigma_N^2} = \Delta\rho \cos\alpha \sin\alpha, \tag{S13}$$

where

$$\frac{\rho_0}{\rho_N} = 1 - \theta_{SH}^2 \frac{2\lambda}{d_N} \tanh\frac{d_N}{2\lambda}, \tag{S14}$$

and

$$\frac{\Delta\rho}{\rho_N} = \theta_{SH}^2 \frac{\frac{2\lambda^2}{\sigma_N d_N}G_r \tanh^2\frac{d_N}{2\lambda}}{1+\frac{2\lambda}{\sigma_N}G_r \coth\frac{d_N}{\lambda}}, \tag{S15}$$

with the electric resistivity $\rho_N = \sigma_N^{-1}$.

The angle dependence agrees perfectly with the experimental results (see Fig. 4 in the main text). We note that a spin transfer at the interface to the FI as well as a finite spin-flip length is necessary for a finite SMR in the present set up.

The measurements of the out-of-plane magnetization angle dependence (see the last panel of Fig. 4 in the main text) are characterized by the angles $\beta$ ($\gamma$) between $\hat{m}$ and $\hat{z}$ when $\hat{x}$ ($\hat{y}$) is the rotation axis. The SMR for this configuration reads

$$\rho = \rho_0 + \Delta\rho \sin^2\beta, \tag{S16}$$

indicating that the resistivity $\rho$ depends on $\beta$, but not on $\gamma$, in agreement with experiments.

The physical quantity for the SMR is the ratio between the change of resistivity and the resistivity

$$\frac{\Delta\rho}{\rho_0} = \frac{\theta_{SH}^2 \frac{2\lambda^2}{\sigma_N d_N} G_r \tanh^2 \frac{d_N}{2\lambda}}{1 - \theta_{SH}^2 \frac{2\lambda}{d_N} \tanh \frac{d_N}{2\lambda}} \approx \theta_{SH}^2 \frac{\frac{2\lambda^2}{\sigma_N d_N} G_r \tanh^2 \frac{d_N}{2\lambda}}{1 + \frac{2\lambda}{\sigma_N} G_r \coth \frac{d_N}{\lambda}}. \quad (S17)$$

According to recent experimental [S5] and theoretical [S4] work the value of the spin-mixing conductance is of the order of $10^{14} \Omega^{-1} m^{-2}$, while the values of the spin diffusion length and the spin Hall angle differ largely [S6]. In Fig. S2, we plot the SMR for a fixed mixing conductance. We observe that the experiments can be explained by a sensible sets of transport parameters ($G_r$, $\lambda$, $\theta_{SH}$) that somewhat differ for the two representative samples.

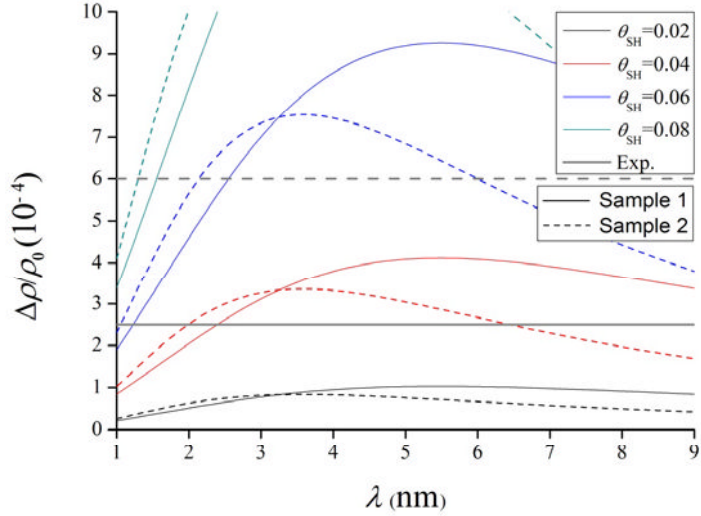

**Figure S2** | Calculated $\Delta\rho/\rho_0$ as a function of $\lambda$ for different spin Hall angles $\theta_{SH}$ with $G_r = 5 \times 10^{14} \Omega^{-1} m^{-2}$. The parameters for the Pt films used here are 12-nm-thick with resistivity $8.6 \times 10^{-7}$ Ωm for the sample 1 (solid curve) and 7-nm-thick with resistivity $4.1 \times 10^{-7}$ Ωm for the sample 2 (dashed curve), respectively. Experimental results are shown as horizontal lines for comparison.

## S3. Analysis of the SMR signal in Pt|Cu|YIG sample

In the Pt|Cu|YIG film, the electric current flows only in the conductive Pt and Cu layers because YIG is an insulator. The electric resistance of the metallic bilayer is roughly described as $R_e = R_{Pt}R_{Cu}/(R_{Pt} + R_{Cu})$ with $R_{Cu}J_{Pt} = R_{Pt}J_{Cu}$, where $R_{Pt(Cu)}$ ($\propto \rho_{Pt(Cu)}/d_{Pt(Cu)}$) and $J_{Pt(Cu)} = [R_{Cu(Pt)}/(R_{Pt(Cu)} + R_{Cu(Pt)})]J_e$ denote the resistance of the Pt(Cu) layer and the electric currents flowing through the Pt(Cu) layer, respectively [S7-S9]. Since the SMR signal is generated only in the Pt layer, the measured total MR signal decreases with increasing the thickness (or the electric conductance) of the Cu layer. In the present system, since $R_{Cu}$ ($\rho_{Cu}$) is one order of magnitude smaller than $R_{Pt}$ ($\rho_{Pt}$), the electric current flows mainly in the Cu layer.

Here, by using the above equivalent circuit model, we estimate the MR amplitude generated in the Pt layer in the Pt|Cu|YIG sample. To extract it from the observed MR in the Pt|Cu|YIG film, we assume the intrinsic MR in the Pt layer as $MR^{int}(d_{Cu}) \approx (1 + R_{Pt}/R_{Cu}) MR(d_{Cu})$. The electric conductances of our samples ($G_e = 1/R_e$) are shown in the inset of Fig. S3a. By using $R_{Pt} = 158$ Ω, we can estimate $R_{Cu}$. The experimentally observed MR and $MR_{trans}$ in the Pt|Cu($d_{Cu}$)|YIG structures are shown in Fig. S3a. Although the measured MR signal decreases with increasing $d_{Cu}$, the values of the intrinsic MR amplitudes generated in the Pt layer, which are estimated by the formula $MR^{int}(d_{Cu}) \approx [1+(\rho_{Pt}/\rho_{Cu})(d_{Cu}/d_{Pt})]MR(d_{Cu})$ and $MR^{int}_{trans}(d_{Cu}) \approx [1+(\rho_{Pt}/\rho_{Cu})(d_{Cu}/d_{Pt})]MR_{trans}(d_{Cu})$, remain unchanged with respect to $d_{Cu}$ (see Fig. S3b). This is consistent with a long spin diffusion length in the Cu layer [S10,S11] which plays a role of an ideal spacer layer for transferring the spin accumulation (spin current) in the Pt layer to YIG through the Cu layer [S12], and strongly supports the proposed SMR mechanism.

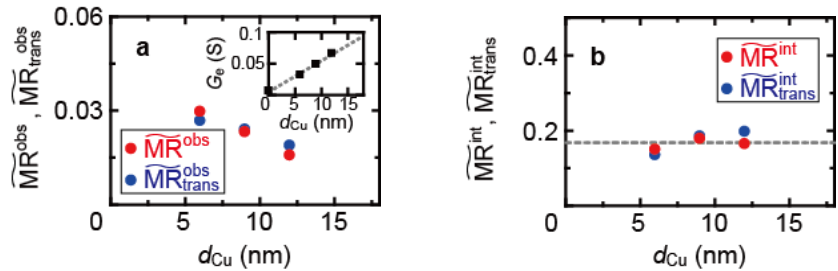

**Figure S3 | Cu thickness $d_{Cu}$ dependence of MR in Pt|Cu|YIG structure. a,** The normalized magnitudes of the observed MR and $MR_{trans}$ in Pt|Cu|YIG system with different $d_{Cu}$; $\widetilde{MR}^{obs} = MR(d_{Cu})/MR(0)$ and $\widetilde{MR}^{obs}_{trans} = MR_{trans}(d_{Cu})/MR_{trans}(0)$. The inset shows the $d_{Cu}$ dependence of the electric conductance in Pt|Cu|YIG; $G_e$. The gray dashed line shows the linear fit to the data. **b,** The normalized magnitudes of the intrinsic MR and $MR_{trans}$ generated in the Pt layer estimated from the equivalent circuit model; $\widetilde{MR}^{int} = MR^{int}(d_{Cu})/MR(0)$ and $\widetilde{MR}^{int}_{trans} = MR^{int}_{trans}(d_{Cu})/MR_{trans}(0)$. The gray dashed line shows the average of the data.